\documentclass[12pt]{article}
\usepackage{amsmath}
\usepackage{graphicx}
\usepackage{enumerate}
\usepackage{natbib}
\usepackage{url} 
\usepackage{authblk}  
\usepackage{amsmath,amssymb,amsthm,amsfonts,natbib,bm,setspace,graphics,graphicx,url,caption,multicol,longtable,lscape,verbatim,enumerate,color,float, diagbox,pdfpages,mathtools, makecell, multirow}

\newcommand{\blind}{0}

\begin{document}

\def\spacingset#1{\renewcommand{\baselinestretch}%
{#1}\small\normalsize} \spacingset{1}


\if1\blind
{
\date{}
  \title{\bf A Scalar-on-Quantile-Function Approach for
Estimating Short-term Health Effects of
Environmental Exposures}
  \author{Yuzi Zhang\thanks{
    The authors gratefully acknowledge \textit{please remember to list all relevant funding sources in the unblinded version}}\hspace{.2cm}\\
    Department of YYY, University of XXX\\
    and \\
    Author 2 \\
    Department of ZZZ, University of WWW}
  \maketitle
} \fi

\if0\blind
{
\date{}
\title{\textbf{A Scalar-on-Quantile-Function Approach for
Estimating Short-term Health Effects of
Environmental Exposures}} 
\author[1]{Yuzi Zhang \thanks{yuzi.zhang@emory.edu}\hspace{.2cm}} \author[1]{Howard H. Chang} \author[2]{Joshua L. Warren} \author[3]{Stefanie T. Ebelt}
\affil[1]{\small Department of Biostatistics and Bioinformatics, Emory University, Atlanta, GA, USA} 
\affil[2]{\small Department of Biostatistics, Yale University, New Haven, CT, USA}
\affil[3]{\small Department of Environmental Health, Emory University, Atlanta, GA, USA}
\maketitle
} \fi

\if1\blind
{
  \bigskip
  \bigskip
  \bigskip
  \begin{center}
    {\LARGE\bf A Scalar-on-Quantile-Function Approach for
Estimating Short-term Health Effects of
Environmental Exposures}
\end{center}
  \medskip
} \fi

\bigskip
\begin{abstract}
Environmental epidemiologic studies routinely utilize aggregate health outcomes to estimate effects of short-term (e.g., daily) exposures that are available at increasingly fine spatial resolutions. However, areal averages are typically used to derive population-level exposure, which cannot capture the spatial variation and individual heterogeneity in exposures that may occur within the spatial and temporal unit of interest (e.g., within day or ZIP code). We propose a general modeling approach to incorporate within-unit exposure heterogeneity in health analyses via exposure quantile functions. Furthermore, by viewing the exposure quantile function as a functional covariate, our approach provides additional flexibility in characterizing associations at different quantile levels. We apply the proposed approach to an analysis of air pollution and emergency department (ED) visits in Atlanta over four years. The analysis utilizes daily ZIP code-level distributions of personal exposures to four traffic-related ambient air pollutants simulated from the Stochastic Human Exposure and Dose Simulator. Our analyses find that effects of carbon monoxide on respiratory and cardiovascular disease ED visits are more pronounced with changes in lower quantiles of the population’s exposure. Software for implement is provided in the R package \texttt{nbRegQF}. 
\end{abstract}

\noindent%
\vfill

\newpage
\spacingset{1.9} 

\section{Introduction}
\label{sec:intro}
Environmental epidemiological studies routinely utilize aggregate health data to assess short-term health effects of environmental exposures. For example, city-wide daily counts of deaths, emergency department (ED) visits, and preterm births have been linked to short-term exposure to air pollution and extreme temperature \citep{alhanti2016ambient, guo2017heat, bekkar2020association, yoo2021association}. The use of aggregate health data represents an ecological design, where the true exposure corresponds to the distribution of individual-level exposure across the at-risk population \citep{richardson2003bayesian, sheppard2003insights}. However, individual environmental exposures cannot be practically measured in large population-based studies. Instead, exposure surrogates that reflect summaries of individual-level exposure (e.g., the mean) are used. This can result in exposure misclassification that leads to biases and incorrect characterization of uncertainties associated with the health effect estimates \citep{dominici2000measurement, richmond2020influence}.

In the context of air pollution, ambient pollutant levels are frequently the exposure of interest so that regulatory policies may be developed. Within short-term time-scales of interest, there can be considerable exposure heterogeneity within the population since (1) air pollutants can exhibit spatial variation within the study region, and (2) individuals spend most of their time in different indoor environments, each with unique outdoor to indoor pollution infiltration characteristics. To address the first challenge, spatial-temporal models have been developed to estimate pollutant concentrations at fine spatial resolutions that can provide complete spatial-temporal coverage \citep{jerrett2001gis, jerrett2005review}. Population-weighting is then used to construct short-term exposure values to link with aggregated health outcome data, over spatial units of interest (e.g., ZIP code). Accounting for spatial exposure variability within temporal studies has been shown to result in larger health effect estimates compared to the use of exposure metrics measured by sparse, stationary monitors, especially for pollutants with high degrees of spatial variability (e.g., traffic-related pollutants) \citep{goldman2010ambient, sarnat2013application}.

In contrast, less work has been done to address exposure heterogeneity due to individual time spent in different microenvironments. With the development of wearable devices, personal exposure to environmental contaminants can be more easily measured \citep{steinle2013quantifying, steinle2015personal, sugg2018temporal}. While the cost is prohibitive for population-based epidemiological studies, findings from small exposure assessment studies have been used to develop probabilistic models to simulate population-level distributions of personal exposures, incorporating time-activity-location patterns and parameters describing the relationships between ambient and indoor environments. When spatially-refined ambient exposure is used as an input, the resulting simulated personal exposure distributions reflect both spatial heterogeneity in ambient concentrations and population heterogeneity in time-activity-location patterns. Examples of probabilistic models include pCNEM and the Stochastic Human Exposure and Dose Simulation (SHEDS) \citep{burke2001population, zidek2005using, zidek2007framework}, as well as a statistical emulator to reduce the computational burden for large-scale epidemiological studies \citep{chang2012time}.

With simulated personal exposures from such probabilistic models, previous studies of aggregate health outcomes have used exposure distribution summary statistics (e.g., daily mean or median) as the covariate in the health model \citep{calder2008relating, chang2012time, sarnat2013application}. However, ecological bias is still present because the daily mean of personal exposures does not fully capture heterogeneity of exposures experienced by the population \citep{sheppard2003insights}. For example, in the special case where personal exposures are normally distributed, inclusion of exposure variance in the health model can result in unbiased estimates \citep{sheppard2003insights, reich2009analysis}. However, the population distribution of environmental exposure is often skewed and poorly approximated by a normal distribution \citep{leiva2008generalized, huang2018multivariate}. Furthermore, all previous methods using population summary statistics implicitly assume that the heath effect of air pollution can be entirely characterized by the summary statistic selected. In other words, health risks only depend on changes in that selected summary statistic. However, health risks may also depend on changes in the expsoure distribution that cannot be reflected by changes in the selected summary statistic. For example, consider a scenario where a short-term increase in air pollution may be more detrimental to individuals who are typically exposed to lower pollution levels. For such scenario, an increase in the lower tail of population-level exposure results in larger increases in the risk of adverse health events, while solely including the population-level mean exposure is insufficient to fully characterize the exposure-response relationship.

In this work, we propose a general modeling framework to incorporate within-unit population-level exposure heterogeneity via exposure quantile functions. Instead of only using unit-level population-average or variance, exposure quantile functions comprehensively summarize the entire within-unit exposure distribution throughout the study. In our framework, the exposure quantile function is viewed as a functional covariate with respect to quantile levels. Therefore, we further allow effects of exposure at different quantile levels to vary. Estimation and inference are carried out under a Bayesian hierarchical modeling framework that also propagates uncertainties associated with the estimation of exposure quantile functions into the health effect estimate.

\section{Motivating Data and Application} \label{s:data}
Ambient air pollution exposure has been identified as a risk factor of various diseases \citep{landrigan2017air, manisalidis2020environmental}, contributing significantly to global disease burden \citep{boogaard2019air}. Here we studied short-term associations between daily ED visits and ambient air pollution exposures in a time-series design \citep{bhaskaran2013time}. We obtained daily counts of ED records during the period January 1st, 1999 - December 31st, 2002 in Atlanta. Counts were also stratified by one on the 40 ZIP code tabulation areas (ZCTAs). We analyzed three causes for ED visits identified using International Classification of Diseases 9th Revision (ICD-9) diagnosis code: (1) respiratory disease (ICD-9 codes: 460-465, 466.0, 477, 466.1, 466.11, 466.19, 480-486, 491-493, 496, 786.07), (2) a subset of respiratory diseases which only includes asthma or wheeze (ICD-9 codes: 493, 786.07), and (3) cardiovascular disease (ICD-9 codes: 410-414, 427-428, 433-437, 440, 443-445, 451-453).

We examined four traffic-related air pollutants: particulate matter with aerodynamic diameter $<2.5$ microns (PM$_{2.5}$), carbon monoxide (CO), nitrogen oxides (NO$_\text{x}$), and elemental carbon (EC), a constituent of PM$_{2.5}$. Separately for each pollutant, population-level distributions of personal exposure were obtained from SHEDS model \citep{burke2001population}. This model is a stochastic simulator producing daily personal exposure at the census tract level \citep{ozkaynak1996particle, jenkins1996personal, dionisio2013development}. To estimate the personal exposure distributions of ambient concentrations of an air pollutant, the SHEDS model first simulates exposures for multiple hypothetical individuals for each census tract that reflect the demographic characteristics (e.g., age, sex, work locations) of the at-risk population using Census data. The amount of time each hypothetical individual spends in various microenvironments is obtained by randomly assigning an activity diary from the US Environmental Protection Agency (EPA)'s Consolidated Human Activity Database based on their demographics. Their daily personal exposure is then computed by summing time-weighted average exposure across all thirteen microenvironments that are categorized into four types (outdoors, vehicle, residential indoors, and non-residential indoors microenvironments). For this analysis, the personal exposure distribution at the census tract level were then aggregated to the ZCTA level. 

For the health analysis, several meteorological variables were obtained from Daymet to account for potential confounding by meteorology \citep{thornton2016daymet}. These include daily minimum temperature, maximum temperature, and dew-point temperature.

\section{Method}
\subsection{Model for count outcome and mean exposure}
We first describe the conventional log-linear model for aggregate outcomes. Let $y_i$ denote the number of events (e.g., ED visits, hospital admissions, deaths) observed for group $i=1,\dots,n$, and let $\boldsymbol{x}_{i} = (x_{i1},\dots,x_{im_i})^T$ denote a vector of exposures (e.g., personal exposures to air pollution) collected from $m_i$ individuals for group $i$. A group can be formed by geographical areas or a time interval. In most applications, the mean of the group-specific exposures are used in an over-dispersed Poisson log-linear regression model such as:
\begin{equation} \label{healthmodel_mean}
  \begin{gathered}
   Y_i|\lambda_i \sim Poisson(\lambda_i), \quad \lambda_i|\xi, \eta_i \sim Gamma(\xi, \exp(\eta_i)), \text{ and} \\
   \eta_i = \alpha \mu_i + \boldsymbol{\gamma}^T\boldsymbol{Z}_i + \epsilon_i,
\end{gathered}  
\end{equation}
where $\mu_i$ denotes the true mean exposure of group $i$ often estimated by the sample average $\sum_{j=1}^{m_i} x_{ij}/m_i$, parameter $\alpha$ represents the effect of the mean exposure, $\boldsymbol{Z}_i$ is a vector of other covariates with regression coefficients $\boldsymbol{\gamma}$ (including an intercept), $\xi$ controls the amount of over-dispersion, and $\epsilon_i$ represents a mean-zero spatial/temporal residual process. Using this parametrization, the log of the expectation of $Y_i$ equals to $\log(\xi) + \eta_i$. Model (\ref{healthmodel_mean}) only depicts the association between expected health outcome count and population-wide mean exposure which ignores exposure heterogeneity among individuals. Therefore, model (\ref{healthmodel_mean}) may be an insufficient framework to characterize effects of exposure on health. For example, model (\ref{healthmodel_mean}) fails to capture health effects associated with changes in exposure distributions having same population-wide means.

\subsection{Model for count outcome and exposure distribution}
To capture effects of the entire exposure distribution and to allow effects to vary by quantile levels, we propose a model treating exposure quantile functions as functional covariates. The proposed scalar-on-quantile-function over-dispersed Poisson log-linear regression model is given as:
\begin{equation} \label{healthmodel}
  \begin{gathered}
   Y_i|\lambda_i \sim Poisson(\lambda_i), \quad \lambda_i|\xi, \eta_i \sim Gamma(\xi, \exp(\eta_i)), \text{ and} \\
   \eta_i = \int_0^1\beta(\tau)Q_i(\tau)d\tau + \boldsymbol{\gamma}^T\boldsymbol{Z}_i + \epsilon_i,
\end{gathered}  
\end{equation}
where $Q_i(\cdot)$ denotes the exposure quantile function of a continuous exposure for group $i$. Note that $\beta(\tau)$ represents the effect of the exposure's $\tau$-th percentile on the mean of the health outcome while other parameters were previously described in model (\ref{healthmodel_mean}).

To flexibly characterize the association between health outcome and the exposure, the coefficient $\beta(\tau)$ is assumed to be a smooth function of quantile levels and is modeled via a finite number of basis functions. Specifically, $\beta(\tau)$ is specified as: 
\begin{equation} \label{betataubasis}
    \beta(\tau) = \sum_{j=0}^{p}K_{j,p}(\tau)\beta_j,
\end{equation}
where $K_{j,p} = \Big(\sqrt{2(p-j)+1}\Big)(1-\tau)^{p-j}\sum_{k=1}^{j}(-1)^{k}{{2p+1-k}\choose{j-k}}{{j}\choose{k}}\tau^{j-k}$ denotes the orthonormal Bernstein polynomials of degree $p$ defined over the interval $[0, 1]$ \citep{bellucci2014explicit}. With this basis expansion, the proposed model in Eqn. (\ref{healthmodel}) can be written as:
\begin{equation} \label{healthmodel1}
\eta_i = \boldsymbol{\beta}^{T} \int_0^1 \boldsymbol{K}(\tau) Q_i(\tau)d\tau + \boldsymbol{\gamma}^T\boldsymbol{Z}_i + \epsilon_i,
\end{equation}
where $\boldsymbol{\beta} = (\beta_0,\dots,\beta_p)^T$ is a vector of basis coefficients and $\boldsymbol{K}(\tau) = (K_{0,p}(\tau),\dots,K_{p,p}(\tau))^T$ is a vector of basis functions. We note that the domain of Bernstein polynomials coincides with the domain of exposure quantile functions, which could facilitate the estimation of $\boldsymbol{\beta}$, compared to B-splines or Gaussian kernels. The Bernstein polynomials are chosen also because they have been shown to accurately approximate various smooth function forms with a small number of basis functions \citep{bellucci2014explicit}. We note one special case where effects of exposure are the same at different quantile levels (i.e., $\beta(\tau)$ is a constant in $\tau$). Then the proposed model (\ref{healthmodel}) will reduce to model (\ref{healthmodel_mean}) because $\int_0^1 Q_i(\tau)d\tau = \mu_i$.

\subsection{Estimation and inference}
\subsubsection{Quantile functions are known}
With known quantile functions, Eqn. (\ref{healthmodel1}) can be reparametrized as a regular scalar-on-scalar model with a covariate vector $\boldsymbol{X}_i^{*}=\int_0^1 \boldsymbol{K}(\tau) Q_i(\tau)d\tau$. For example, in \cite{reich2009analysis}, $Q_{i}(\tau)$ is assumed to be Normal quantile functions. An efficient fully Bayesian inference procedure is available for the over-dispersed Poisson regression by introducing latent Polya-Gamma random variables \citep{polson2013bayesian, neelon2019bayesian}. Using the Polya-Gamma method, coefficients $\boldsymbol{\beta}$ and $\boldsymbol{\gamma}$ can be estimated via Markov Chain Monte Carlo (MCMC) using Gibbs sampling. With independent normal priors introduced for coefficients $\boldsymbol{\beta}$ and $\boldsymbol{\gamma}$, the full conditional distribution for those regression coefficients is a multivariate normal distribution.


\subsubsection{Quantile functions are unknown}
For the more realistic case of unknown quantile functions, we propose a two-stage Bayesian estimation procedure. In the first stage, quantile function $Q_i(\tau)$ for each group $i$ is again modeled using basis expansion and estimated from individual-level exposures (e.g., SHEDS simulations in our application). In the second stage, the health model (\ref{healthmodel}) is fitted with estimated quantile functions while accounting for the statistical uncertainties associated with the first-stage estimation.

Following previous semiparametric Bayesian approaches for modeling quantile processes for continuous variables \citep{reich2012spatiotemporal}, the quantile function for exposures in group $i$ is expanded as:
\begin{equation}
    Q_i(\tau) = \theta_{0,i} + \sum_{l=1}^{L}B_l(\tau)\theta_{l,i},
\end{equation}
where $B_l(\tau)$ is the $l$-th basis function, and $\theta_{l,i}$ are basis coefficients. Choices of basis function $B_l(\tau)$ include piecewise Gaussian or piecewise Gamma functions, and their expressions are provided in supplementary materials. Both choices permit us to flexibly characterize the potentially skewed distribution of exposures. However, for exposures that are strictly positive (e.g., ambient air pollution concentration), piecewise Gamma functions are recommended. With the use of piecewise Gaussian or Gamma functions, $\theta_{0,i}$ represents the median of the $i$-th group exposure distribution and $\theta_{l,i}$ for $l=1\dots,L$ characterize the shape of the distribution. The quantile function uniquely defines the density. Let $x_{ij}$ denote the exposure level measured for the $j$-th individual within the $i$-th group. When $x_{ij}$ is assumed to follow a distribution corresponding to a quantile function $Q_i(\tau)$, the likelihood for a set of individual-level exposures is given by
\begin{equation}
L(\boldsymbol{\theta}_{0,\boldsymbol{\cdot}},\{\boldsymbol{\theta}_{l,\boldsymbol{\cdot}}\}_{l=1}^{L};\{\boldsymbol{x}_i\}_{i=1}^{n}) = \prod_{i=1}^{n} \prod_{j=1}^{m_i}\Big[\frac{d Q_i(\tau)}{d \tau}\Big]^{-1} \biggr\rvert_{\tau^*:Q_{i}(\tau^*) = x_{ij}},  
\end{equation}
where $\boldsymbol{\theta}_{0,\boldsymbol{\cdot}} = (\theta_{0,1}, \dots,\theta_{0,n})^T$, $\boldsymbol{\theta}_{ l,\boldsymbol{\cdot}} = (\theta_{l,1}, \dots,\theta_{l,n})^T$, and $\boldsymbol{x}_{i} = (x_{i1}, \dots, x_{im_i})^T$ is a vector of personal exposures collected for group $i$ of $m_i$ individuals. It is important to note that quantile functions have to be non-decreasing. To ensure this property, $\theta_{l,i} > 0$ should hold for any $i$ and $l$ \citep{reich2012spatiotemporal}. In the first-stage estimation, this constraint is imposed by introducing an unconstrained latent variable $\theta_{l,i}^{*}$. Specifically, $\theta_{l,i} = \texttt{max}\{\theta_{l,i}^{*}, \nu \}$, where $\nu$ is a small constant (e.g., 0.01). For spatial or time-series data, one can easily introduce spatial or temporal dependence for $\boldsymbol{\theta}_{0,\boldsymbol{\cdot}}$ and $\boldsymbol{\theta}_{l,\boldsymbol{\cdot}}^{*} = (\theta_{l,1}^*,\dots,\theta_{l,n}^{*})^T$ in the Bayesian hierarchical modeling framework to allow quantile processes to vary by time or locations. Estimation of all model parameters is carried out via MCMC algorithms. 

In the second-stage, the health model is fitted while accounting for the uncertainties associated with estimating exposure quantile functions. When the exposure quantile function is expanded using basis functions, the health model in Eqn. (\ref{healthmodel1}) becomes:
\begin{equation} \label{healthmodel2}
 \eta_i = \boldsymbol{\beta}^{T} \Big[ \int_0^1 \boldsymbol{K}(\tau) \boldsymbol{B}(\tau)^{T} d\tau \Big] \boldsymbol{\theta}_{\boldsymbol{\cdot}, i}  + \boldsymbol{\gamma}^T\boldsymbol{Z}_i + \epsilon_i,
\end{equation}
where $\boldsymbol{B}(\tau) = (1, B_{1}(\tau), \dots, B_{L}(\tau))^{T}$ and $\boldsymbol{\theta}_{\boldsymbol{\cdot},i} = (\theta_{0, i}, \theta_{1,i}, \dots, \theta_{L,i})^{T}$. Since $\boldsymbol{B}(\tau)$ is pre-specified, uncertainties of estimated $\boldsymbol{\theta}_{\boldsymbol{\cdot},i}$, contribute to uncertainties of the estimation of quantile functions. 

To appropriately propagate uncertainties resulting from the first-stage estimation, we consider an approach which is commonly used in environmental health studies for incorporating uncertainties in estimated exposures \citep{carroll2006measurement, lee2017rigorous}. Specifically, a multivariate normal (MVN) prior is assumed for $\boldsymbol{\theta}_{\boldsymbol{\cdot},i}$ with mean and variance-covariance matrix computed from its posterior predictive distribution obtained from the first-stage estimation. Similarly to the case in which quantile functions are known, we view $\boldsymbol{X}_{i}^{*} = \big[ \int_0^1 \boldsymbol{K}(\tau) \boldsymbol{B}(\tau)^{T} d\tau \big] \boldsymbol{\theta}_{ \boldsymbol{\cdot},i}$ as a random covariate vector. With MVN prior assumed for $\boldsymbol{\theta}_{\boldsymbol{\cdot},i}$, the prior of $\boldsymbol{X}_{i}^{*}$ is also MVN. It is worth noting that posterior distributions of $\boldsymbol{\theta}_{\boldsymbol{\cdot},i}$ are correlated across groups when quantile processes are assumed to vary by groups. As the number of groups increases, this approach becomes computational expensive since the MCMC algorithm requires sampling from a high-dimensional MVN distribution. In the simulation study and real data analysis, we ignore the correlation between groups to facilitate the computation. We find that this does not meaningfully impact inference for the health effect association, as also recently demonstrated in \cite{comess2022bayesian}. As in the case of known quantile functions, the estimation of coefficients $\boldsymbol{\beta}$ and $\boldsymbol{\gamma}$ is carried out by Gibbs sampling with normal priors specified for those coefficients using the Polya-Gamma method. Details of MCMC algorithms for the estimation of quantile functions, and the estimation with known and unknown quantile functions can be found in the supplementary materials.

\section{Simulation Studies} \label{s:sim}
We conducted simulation studies to examine the impact of not accounting for exposure heterogeneity motivated by our application and evaluated the performance of the two-stage estimation procedure. A variety of associations between exposures and outcomes are considered by specifying different forms of coefficient regression function $\beta(\tau)$. 

The simulation study assumes that health outcome is collected over $n=1000$ time points and exposure quantile functions are temporally correlated. The temporal dependence was introduced by defining a first-order Gaussian Markov random field process for the unconstrained latent basis coefficients. The true health model and exposure quantile function are:
\begin{equation*}
    \begin{gathered}
   Y_i|\lambda_i \sim Poisson(\lambda_i), \quad \lambda_i|\xi, \eta_i \sim Gamma(\xi, \exp(\eta_i)), \\
   \eta_i = \beta_{0} + \int_0^1\beta(\tau)Q_i(\tau)d\tau, \; i = 1, \dots, 1000, \\
   Q_{i}(\tau) = \theta_{0,i} + \sum_{l=1}^{4}B_{l}(\tau)\theta_{l,i},\\
   (\theta_{0,1}, \dots, \theta_{0,1000})^T \sim MVN(\theta_{0}, \Sigma_0), \; \Sigma_0 = \sigma_{0}^{2}(D_w - \rho_{0}W)^{-1}, \\
    (\theta_{l,1}^{*}, \dots, \theta_{l,1000}^{*})^T \sim MVN(\theta_{l}, \Sigma_1), \; \Sigma_1 = \sigma_{1}^{2}(D_w - \rho_{1}W)^{-1} \; \text{for } l = 1,\dots, 4, \theta_{l,i} = \texttt{max}\{\theta_{l,i}^{*}, \nu=0.01\}, 
    \end{gathered}
\end{equation*}
where $B_l(\tau)$ are piecewise Gamma functions with 4 basis functions; $W$ is a symmetric matrix with the $(i, i')$ entry equal to 1 if time point $i$ and $i'$ are adjacent, and 0 otherwise; and $D_{w}$ is a diagonal matrix with the $i$-th diagonal element equal to the row sum of $i$-th row of $W$. For each time point $i$, individual exposure data with a sample size of 100 were generated using the quantile function $Q_{i}(\tau)$. To mimic the right-skewed distribution of PM$_{2.5}$ observed in the motivating data, we set $\theta_0 = 7.2$, $\theta_l = 0.9$ for $l=1,\dots, 4$, and parameters controlling temporal correlations of quantile processes $(\sigma_0^2, \rho_0, \sigma_1^2, \rho_1)$ equal to $(1, 0.9, 0.02, 0.9)$. 



\begin{figure}[!htbp]
    \centering
    \includegraphics[width = 0.9\textwidth]{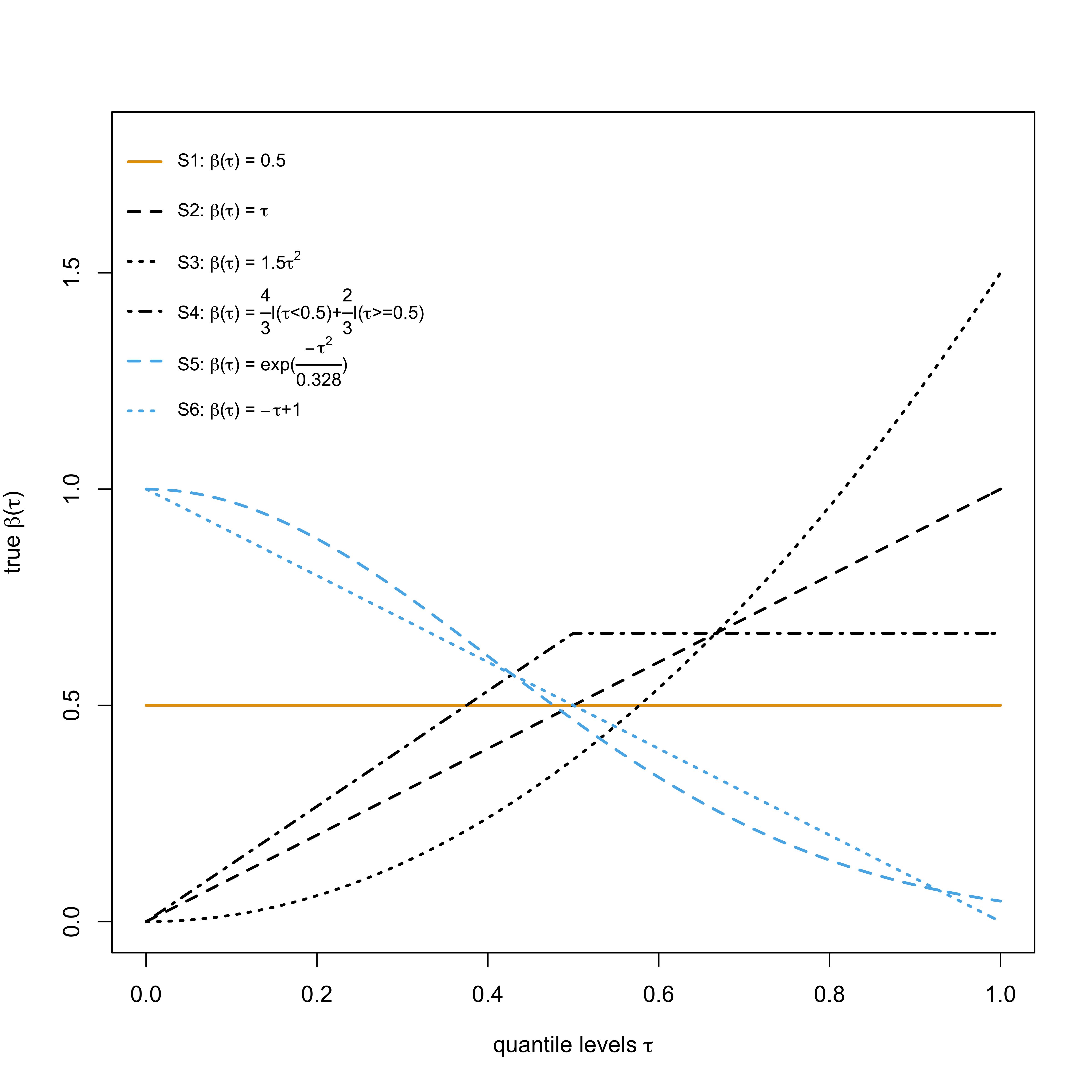}
    \caption{True coefficient regression functions $\beta(\tau)$ specified for the simulation study}
    \label{Fig:truebetatau}
\end{figure}

We examined six coefficient regression functions displayed in Figure \ref{Fig:truebetatau}: (1) $\beta(\tau) = 0.5$, (2) $\beta(\tau) = \tau$, (3) $\beta({\tau}) = 1.5\tau^2$, (4) $\beta(\tau) = \frac{4}{3}\tau I(\tau < 0.5) + \frac{2}{3} I(\tau \ge 0.5)$, (5) $\beta(\tau) = \exp(-\frac{\tau^2}{0.328})$, and (6) $\beta(\tau) = -\tau + 1$. These six functions include three types of effects (constant, increasing, and decreasing). We note that $\int_{0}^{1} \beta(\tau) d\tau$ equals to 0.5 under all scenarios, and this quantity is interpreted as the effect of exposures associated with the exposure distribution shifted to the right by one unit (i.e., the mean is increased by one unit). Based on simulated quantile functions $Q_i(\tau)$, 100 health datasets were generated for each of different forms of $\beta(\tau)$ while fixing $\beta_0 = -3.5$. The conventional ``mean" model using the mean of individual exposures at each time point as the covariate was also fitted for the comparison.

Three quantities were used for performance evaluation: (a) the effect associated with the exposure distribution shifted to the right by one unit, which is measured by $\int \beta(\tau) d\tau$ for the proposed model and parameter $\alpha$ for the mean model; (b) predictive values of the exposure which is measured by $\int_{0}^{1}\beta(\tau)Q_i(\tau)d\tau$ and $\alpha \mu_i$ for the proposed and mean models, respectively; (c) the total number of exposure-attributable events, which is computed as $\sum_{i=1}^{n}\xi\big\{ \exp \big[\beta_0 + \int_{0}^{1}\beta(\tau)Q_i(\tau)d\tau \big] - \exp(\beta_0)\big \}$ and $\sum_{i=1}^{n} \xi\big[\exp(\beta_0 + \alpha \mu_i) - \exp(\beta_0)\big]$ for the proposed and mean models, respectively. For these quantities, their relative bias, mean squared error (MSE), and 95\% coverage probability (CP) were computed. 


For $\int_{0}^{1} \beta(\tau) d\tau$ and the total number of exposure-attributable events, their relative bias, MSE, and CP were computed by averaging over simulations. However, for predictive values, we averaged over both simulations and time points. Specifically, when exposure quantile functions were estimated, the MSE of predictive values of the exposure was calculated as:
\begin{gather}
   \frac{1}{n \times D} \sum_{i=1}^{n} \sum_{d=1}^{D} \Big[ \int_0^1 \hat{\beta}^{(d)}(\tau)\hat{Q}_{i}^{(d)}(\tau) d\tau - \int_0^1 \beta(\tau)Q_i(\tau) d\tau \Big ] ^2  \label{quan_MSE}  
\end{gather}
for the proposed model, where subscript $d$ is the index for the simulation with $n=1000$ and $D=100$; $\hat{\beta}^{(d)}(\tau) = \hat{\boldsymbol{\beta}^{(d)}}^{T}\boldsymbol{K}(\tau)$, $\hat{\boldsymbol{\beta}^{(d)}}$ is the vector of estimated basis coefficients defined in Eqn. (\ref{healthmodel1}); $\hat{Q}_i(\tau)^{(d)} = \boldsymbol{B}(\tau)^T \hat{\boldsymbol{\theta}}^{(d)}_{\boldsymbol{\cdot}, i}$, $\hat{\boldsymbol{\theta}}^{(d)}_{\boldsymbol{\cdot}, i}$ are estimated basis coefficients defined in Eqn. (\ref{healthmodel2}). The computation of the relative bias and CP follows a similar manner. 

Additionally, the estimation performance of $\beta(\tau)$ was examined with respect to its bias, MSE, and CP by averaging over simulations and quantile levels $\tau_j \in \{0, 0.01, \dots, 0.99, 1\}$. For example, MSE of $\beta(\tau)$ is $1/(J \times D) \sum_{j=1}^{J} \sum_{d=1}^{D} \big[ \hat{\boldsymbol{\beta}^{(d)}}^{T}\boldsymbol{K}(\tau_{j}) - \beta(\tau_{j}) \big]^2$. Relative bias was not computed, since $\beta(\tau)$ can take value of 0. 

In this simulation study, we modeled $\beta(\tau)$ using orthonormal Bernstein polynomials of degree two (i.e., three basis functions) and specified vague normal priors N(0,100) for the corresponding basis coefficients. The over-dispersion parameter included in the health model was updated using Metropolis-Hastings (MH) algorithms with an uniform prior. For the estimation of exposure quantile functions, we used four piecewise Gamma functions to expand quantile functions. Coefficients $\theta_{0,i}$ and $\theta_{l,i}^*$ were assigned N(0, 100) priors and were updated using MH algorithms; $\sigma_{0}^2$ and $\sigma_{1}^2$ were given InvGamma(0.1, 0.1) priors and were updated using Gibbs sampling; discrete priors (i.e., 1000 equally spaced values between 0 and 1) were assigned for $\rho_0$ and $\rho_1$ which were updated using MH algorithms. We generated 10,000 MCMC samples and discarded the first 5,000 samples as burn-in when estimating quantile functions; while 5,000 samples were generated and the first 2,500 samples were discarded as burn-in when estimating health effects regardless of using true or estimated quantile functions.  

Simulation results from using different exposure covariates (true mean exposures, true exposure quantile functions, and estimated quantile functions) are summarized in Table \ref{t:simres}. We first focus on the case where exposure quantile functions and mean exposures are assumed to be known. Under the scenario where $\beta(\tau)$ is a constant (i.e., S1), results from using quantile functions and mean of exposures are similar. Since $\beta(\tau)$ was modeled using the basis expansion, a larger MSE was observed for the proposed model. For other scenarios, the proposed model resulted in empirically unbiased estimates of health effects. However, we observed biased estimates of health effects when using mean exposures. Specifically, we found positive biases when the upper tail of exposure distributions has larger effects (e.g., S2 and S3), and negative biases when the effects decrease as the quantile level increases (e.g., S5 and S6). The reverse pattern was observed for the total number of events attributed to exposures and predictive values of exposures. In S4 where the health effect first increases and then becomes a constant, the model using mean exposure happened to lead to a nearly unbiased estimate with conservative 95\% credible intervals (CIs) for $\int_{0}^{1} \beta(\tau) d\tau$. However, this model failed to characterize effects of exposures as indicated by the larger MSE and severe under-coverage associated with predictive values of exposures. It is also worth noting that the proposed model performed slightly worse in this scenario, for example, the MSE of $\int_{0}^{1} \beta(\tau) d\tau$ was higher than S2, S3, and S5, likely because $\beta(\tau)$ in S4 is approximated less well by the basis expansion. 

\begin{singlespace}
\begin{table}[!htbp]
\footnotesize
 \begin{center}
  \caption{Simulation results using different exposure covariates}
\label{t:simres}
\setlength\tabcolsep{2pt}
\renewcommand{\arraystretch}{1.5}
\begin{tabular}[1.2\textwidth]{@{}cc ccc l ccc l ccc l ccc@{}}
\hline \hline 
 & & \multicolumn{3}{c}{$\int_{0}^{1}\beta(\tau)d\tau$ $^b$} & & \multicolumn{3}{c}{$\beta(\tau)$} & & \multicolumn{3}{c}{\makecell{predictive values \\ of exposures}} & &  \multicolumn{3}{c}{\makecell{exposure-attributable \\ events}} \vspace{2pt} \\  \cline{3-5} \cline{7-9} \cline{11-13} \cline{15-17} \vspace*{3pt}
Scenario & Covariate $^a$ & \makecell{relative \\ bias} & \makecell{relative\\MSE $^c$} & \makecell{CP \\(\%)} && bias & \makecell{MSE} & \makecell{CP \\ (\%)} && \makecell{relative \\ bias}  & \makecell{relative \\ MSE$^c$} & \makecell{CP \\ (\%)} && \makecell{relative \\ bias} & \makecell{relative \\ MSE$^c$} & \makecell{CP \\ (\%)} \\
\hline 
\multirow{3}{*}{S1} & mean & 0.004 & 1.00 & 94 && - & - & - && 0.004 & 1.000 & 94.00 && 0.001 & 1.000 & 95 \vspace{2pt} \\
&  quantile & 0.005 & 1.10 & 96 && 0.002 & 0.016 & 93.70 && -0.002 & 3.889 & 93.39 && 0.000 & 1.069 & 95 \\ 
&  \makecell{quantile \\ with errors} & 0.011 & 1.47 & 92 && 0.004 & 0.031 & 95.48 && 0.004 & 6.475 & 96.41 && 0.001 & 1.101 & 94 \vspace{2pt} \\ \hline 
\multirow{3}{*}{S2} & mean & 0.009 & 1.00 & 89 && - & - & - && -0.122 & 1.000 & 0.01 && -0.008 & 1.000 & 83\\
& quantile  & -0.002 & 0.80 & 96 && -0.001 & 0.015 & 92.17 && -0.002 & 0.052 & 93.66 && -0.000 & 0.610 & 93 \vspace{2pt} \\ 
&  \makecell{quantile \\ with errors}  & 0.010 & 1.07 & 91 && 0.003 & 0.032 & 93.13 && 0.010 & 0.108 & 95.50 && 0.000 & 0.620 & 93 \vspace{2pt} \\ \hline 
\multirow{3}{*}{S3} & mean & 0.019 & 1.00 & 75 && - & - & - && -0.174 & 1.000 & 0.00 && -0.010 & 1.000 & 82\\
& quantile  & -0.001 & 0.39 & 97 && -0.000 & 0.013 & 92.92 && -0.003 & 0.023 & 91.74 && 0.000 & 0.408 & 96 \vspace{2pt} \\
&  \makecell{quantile \\ with errors}  & 0.014 & 0.74 & 90 && 0.005 & 0.036 & 90.47 && 0.017 & 0.056 & 94.12 && 0.001 & 0.398 & 96 \vspace{2pt} \\ \hline 
\multirow{3}{*}{S4} & mean & 0.003 & 1.00 & 100 && - & - & - && -0.081 & 1.000 & 0.04 && -0.006 & 1.000 & 89 \\
& quantile & -0.001 & 1.01 & 98 && -0.001 & 0.014 & 93.61 && 0.003 & 0.127 & 93.56 && 0.000 & 0.751 & 95 \vspace{2pt} \\
&  \makecell{quantile \\ with errors} & 0.008 & 1.38 & 98 && 0.001 & 0.039 & 90.57 && 0.015 & 0.302 & 94.82 && 0.001 & 0.759 & 94 \vspace{2pt} \\ \hline 
\multirow{3}{*}{S5} & mean & -0.010 & 1.00 & 95 && - & - & - && 0.182 & 1.000 & 0.00 && 0.017 & 1.000 & 77 \\
& quantile & 0.002 & 0.80 & 95 && 0.001 & 0.018 & 95.90 && -0.003 & 0.057 & 95.73 && 0.000 & 0.361 & 95\vspace{2pt} \\
&  \makecell{quantile \\ with errors} & 0.003 & 0.85 & 97 && 0.000 & 0.033 & 97.12 && -0.002 & 0.107 & 96.02 && -0.000 & 0.381 & 95 \vspace{2pt} \\ \hline 
\multirow{3}{*}{S6} & mean & -0.011 & 1.00 & 94 && - & - & - && 0.163 & 1.000 & 0.02 && 0.014 & 1.000 & 80 \\
& quantile & 0.000 & 0.75 & 95 && 0.000 & 0.016 & 95.74 && -0.001 & 0.075 & 95.67 && -0.001 & 0.537 & 95 \\
&  \makecell{quantile \\ with errors} & 0.002 & 0.83 & 96 && 0.000 & 0.042 & 94.42 && -0.003 & 0.127 & 95.89 && -0.002 & 0.586 & 93 \\
\hline \hline
\end{tabular}    
\renewcommand{\arraystretch}{1}
 \end{center} \vspace*{-10pt}
{$^a$ \footnotesize mean = true mean exposures, quantile = true exposure quantile functions, quantile with errors = estimated exposure quantile functions.} \\ 
{$^b$ \footnotesize For the model using mean exposure as the exposure covariate, the parameter $\alpha$ was reported.} \\ 
{$^c$ \footnotesize Relative MSE was computed by treating the model using the true mean exposure as the reference.}
\end{table}
\end{singlespace}

We also evaluated the decision to select the proposed model over the conventional ``mean" exposure model informed by the widely available information criterion (WAIC) \citep{watanabe2010asymptotic}. Among 100 simulations, the proportion of simulations in which the WAIC favored the mean model is 82\% under S1 where the mean model is a valid and better choice. While for other scenarios, the WAIC favored the proposed model over $95\%$ of the time. 

\begin{figure}[!htbp]
    \centering
    \includegraphics[width = \textwidth]{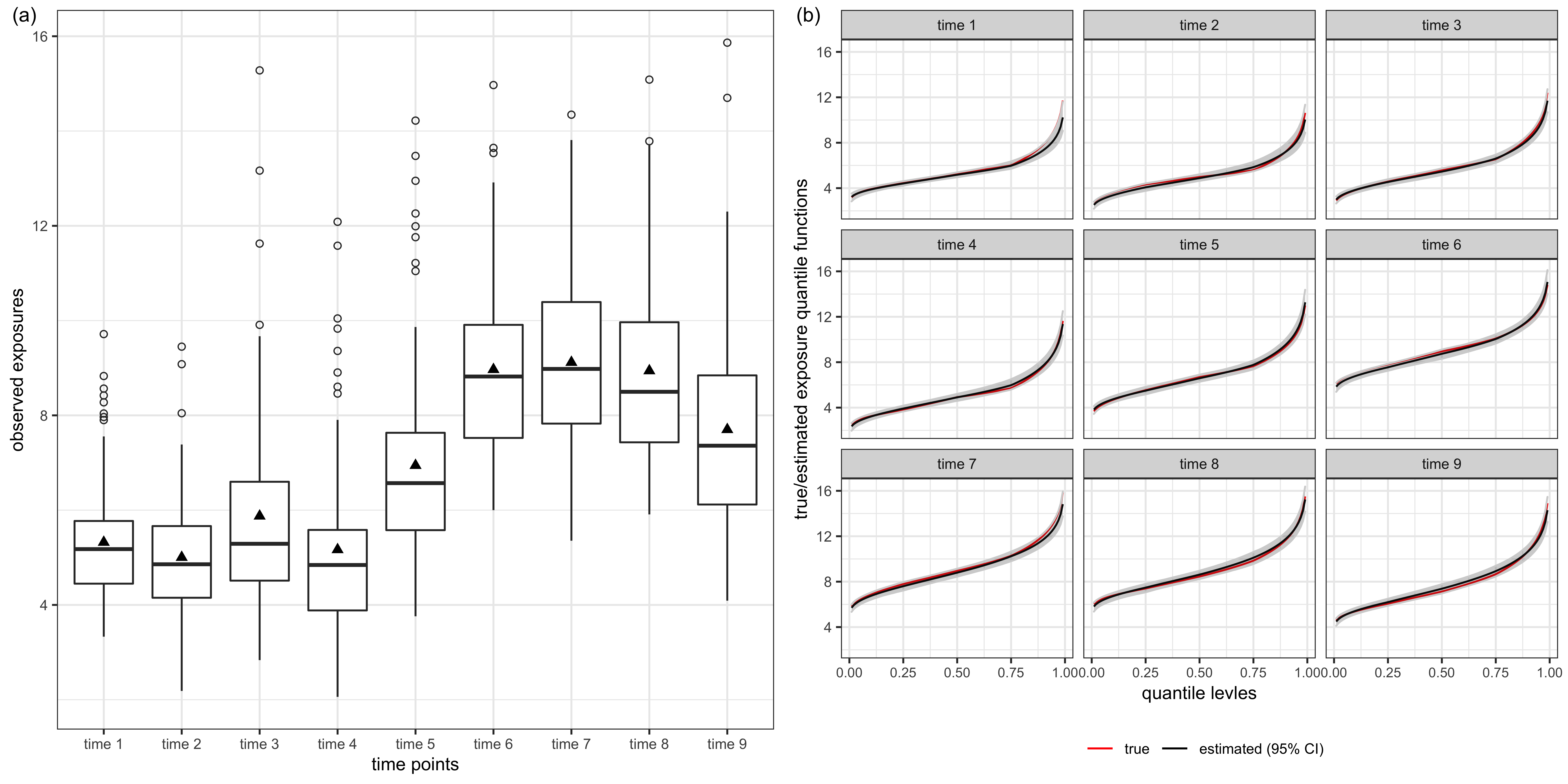}
    \caption{(a) boxplots of exposures observed on nine consecutive days in the simulated data, (b) true/estimated exposure quantile functions for nine consecutive days}
    \label{fig:sim_quan}
\end{figure}

When exposure quantile functions were estimated from the simulated dataset containing individual-level exposures, we applied the two-stage estimation procedure to account for uncertainties of estimating exposure quantile functions. Figure~\ref{fig:sim_quan}(a) shows simulated exposure data over nine consecutive days where exposure distributions are right skewed and change over time smoothly. Estimated quantile functions traced the truth well as shown in Figure~\ref{fig:sim_quan}(b). Compared to the case where true quantile functions were used, MSE of those four quantities were increased (Table \ref{t:simres}). However, across all scenarios, coverage probabilities still achieved or were close to the nominal level. We found that without accounting for estimation uncertainties by using the posterior means of the estimated quantile functions can lead to bias and under-coverage (results are not shown).

\section{Real Data Analysis}
In this section, we analyzed the motivating data introduced in Section~\ref{s:data}  using both the proposed scalar-on-quantile-function model and the conventional model using average concentrations of air pollutants as the exposure. 

\subsection{Estimation of daily exposure quantile functions}
\begin{figure}[!htbp]
    \centering
    \includegraphics[width = \textwidth]{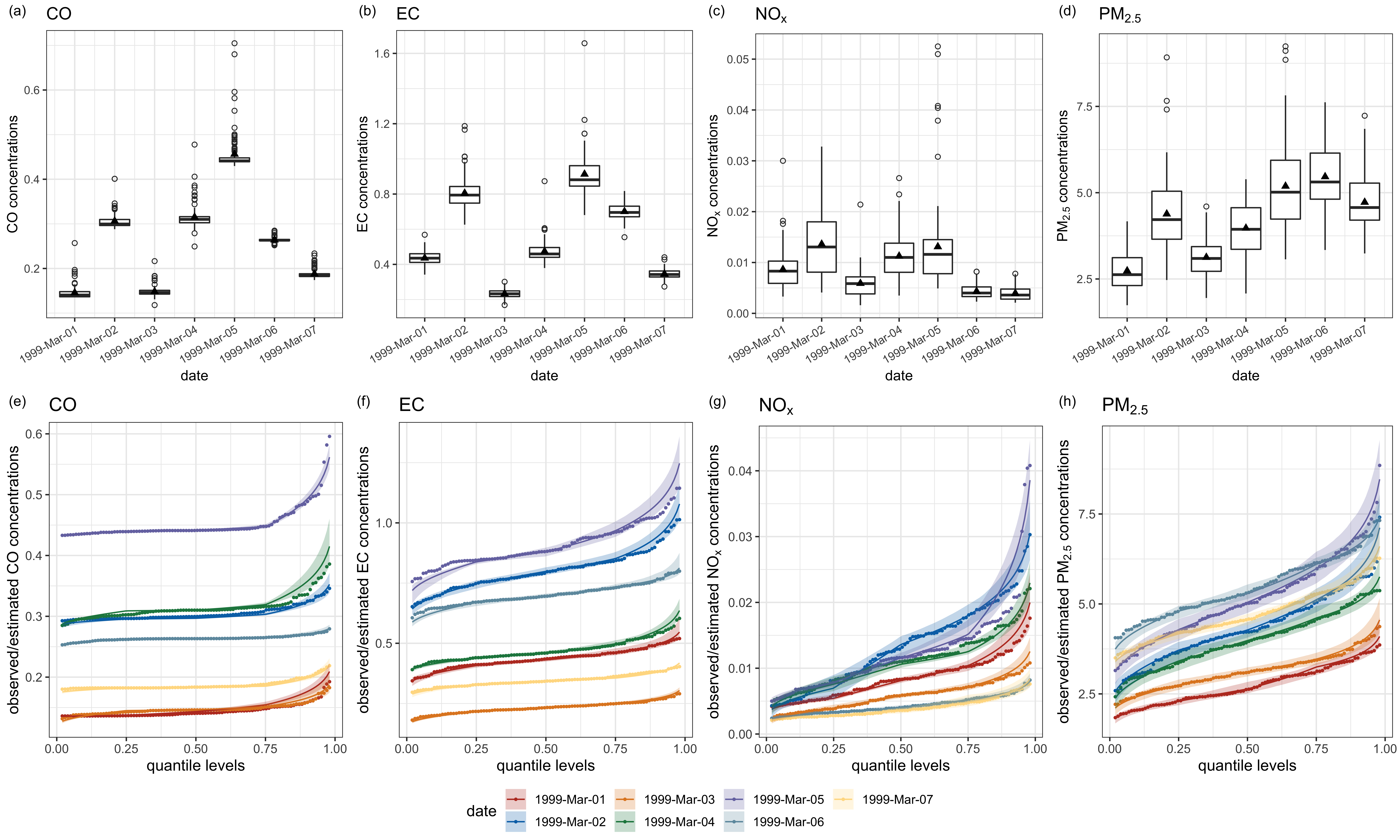}
\caption{(a)-(d) Boxplots of concentrations of air pollutants obtained from SHEDS at ZCTA 30032 on seven representative days, (e)-(h) empirical/estimated quantile functions with 95\% credible intervals of air pollutants obtained from SHEDS at ZCTA 30032 on seven representative days (empirical quantile functions are denoted by solid points).}
    \label{f:box_quan_SHEDS}
\end{figure}

Distributions of four air pollutants from SHEDS at one representative ZCTA are presented in Figures \ref{f:box_quan_SHEDS}(a)-(d). We observed different degrees of skewness in exposure distributions for all four air pollutants. We noted that distributions of CO and EC are very different across days compared to NO$_\text{x}$ and PM$_{2.5}$. Hence, for estimating quantile functions, quantile functions of CO and EC were assumed to be independent across days and ZCTAs, while temporal correlations were introduced for quantile functions of NO$_\text{x}$ and PM$_{2.5}$. In this analysis, we used the same priors, the number of MCMC iterations, and burn-in as in Section~\ref{s:sim}. Figures \ref{f:box_quan_SHEDS}(e)-(h) shows the corresponding empirical quantile and estimated quantile functions from using four piecewise Gamma functions. Overall, our use of basis functions sufficiently capture exposure distributions with larger uncertainties in lower and upper tails as expected. 

\subsection{Estimation of health effects for ED visits}
Associations between ED visits and same-day air pollution concentrations were examined using the proposed model with estimated quantile functions and uncertainty propagation, and the conventional model using sample mean of individual exposures at the ZCTA level. Following the previous health analysis \citep{sarnat2013application}, the following confounders were controlled for in all models: non-linear effect of year-specific temporal trends using day of year, non-linear effect of same-day dew-point temperature, indicators of day of week, an indicator of federal holidays, non-linear effect of 3-day moving average of minimum temperature (maximum temperature was controlled for cardiovascular disease ED visits). All non-linear effects were modeled with natural cubic splines with four degrees of freedom. Exchangeable ZCTA-specific random intercepts were included for all models. We varied the number of basis functions used for modeling $\beta(\tau)$ from two to three, and results from the model with lower WAIC are reported. The same priors as in Section~\ref{s:sim} were introduced for basis coefficients and the over-dispersion parameter; the variance parameter of spatial random effects were assumed to have InvGamma(0.1, 0.1) prior. We ran for 7,500 iterations, the first 3,500 being discarded as burn-in. 

\begin{figure}[!htbp]
    \centering
    \includegraphics[width = \textwidth]{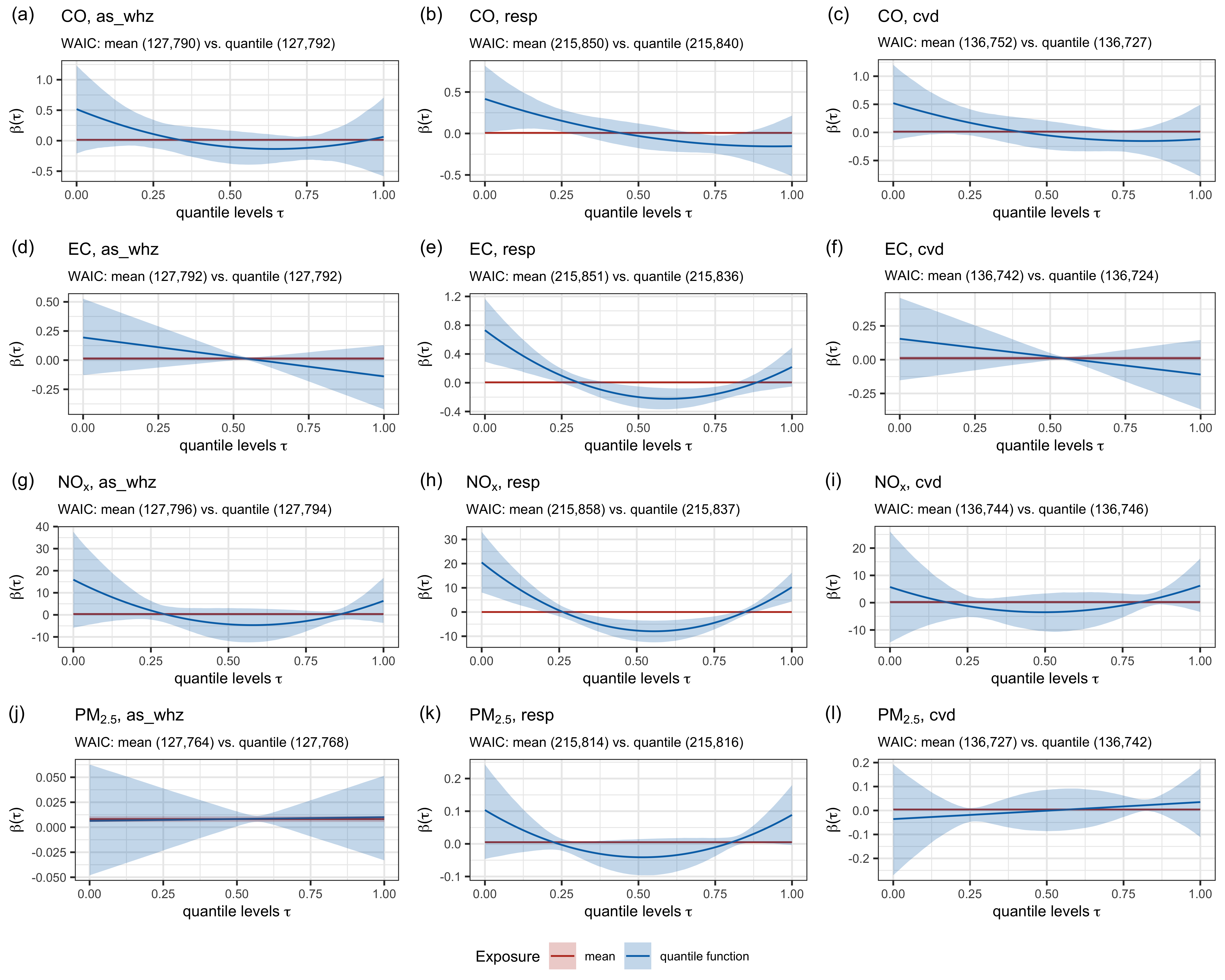}
    \caption{Estimates of $\beta(\tau)$ with 95\% credible intervals from analyzing the motivating data (as\_whz = asthma or wheeze ED visits, resp = respiratory disease ED visits, cvd = cardiovascular disease ED visits).}
    \label{fig:SHEDSbetatau}
\end{figure}

Figure \ref{fig:SHEDSbetatau} plots the estimated $\beta(\tau)$ with 95\% CIs for different combinations of air pollutants and causes of ED visits. We observed that the WAIC clearly favors the proposed model under some cases (e.g., panels (b), (c), (e), (f), and (h) of Figure \ref{fig:SHEDSbetatau}). For example, Figure \ref{fig:SHEDSbetatau}(e) shows that lower and upper tails of the distribution of EC concentrations have larger effects on respiratory disease ED visits. The resulting percent (\%) increase in risk associated with one unit shift to right for the distribution of EC concentrations is estimated to be 2.58 (95\% CI: 0.82 - 4.31) and 0.56 (95\% CI: -0.14 - 1.26) using the proposed and the conventional models, respectively. In some cases, the WAIC indicates that the model utilizing quantile functions and the model using sample mean of exposures fitted the data equally. For example, Figure \ref{fig:SHEDSbetatau}(j) shows that estimates of health effects of PM$_{2.5}$ on asthma or wheeze ED visits basically remain the same across quantile levels, matching the estimate obtained from the mean model. Estimates of percent increase in risk for the rest of combinations are presented in Table S1. 

\begin{figure}[!htbp]
    \centering
    \includegraphics[width = \textwidth]{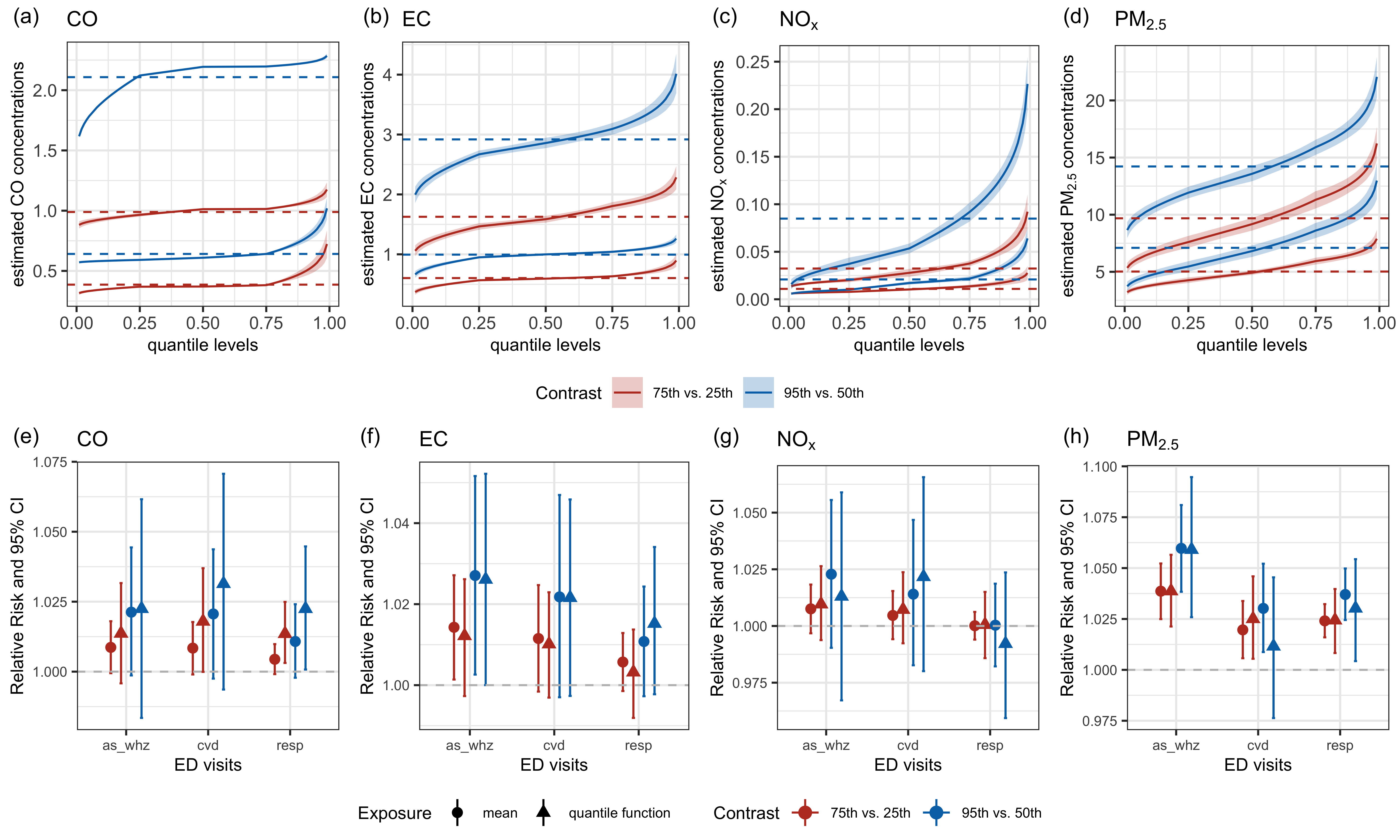}
    \caption{Short-term associations between ED visits associated with two observed exposure distributions defined by the exposure median (75$^{\text{th}}$ versus 25$^{\text{th}}$, and 95$^{\text{th}}$ versus 50$^{\text{th}}$ percentiles of the distribution of estimated medians across all ZCTAs and days). Results obtained from models using the mean concentrations as the exposure are also shown. as\_whz = asthma or wheeze, resp = respiratory disease, cvd = cardiovascular disease.}
    \label{fig:SHEDSRR}
\end{figure}

Figures \ref{fig:SHEDSRR}(a)-(d) display estimated quantile functions where their corresponding exposure medians are at the 25$^{\text{th}}$, 50$^{\text{th}}$, 75$^{\text{th}}$, and 95$^{\text{th}}$ percentiles of the estimated medians across all ZCTAs and days. Dashed horizontal lines mark mean exposure calculated from SHEDS data used for estimating the quantile functions. These quantile functions were selected to represent different exposure contrasts across ZCTAs and days. Estimates of relative risks associated with changes in selected estimated quantile functions from the proposed and the conventional models are shown in Figures \ref{fig:SHEDSRR}(e)-(h). Two contrasts were selected to represent different exposure effects: (1) comparing quantile functions with medians at the 75$^{\text{th}}$ and 25$^{\text{th}}$ percentiles to represent typical exposure contrast, and (2) comparing quantile functions with medians at the 95$^{\text{th}}$ and 50$^{\text{th}}$ percentiles to represent more extreme exposure effects.  

\begin{figure}[!htbp]
    \centering
    \includegraphics[width = \textwidth]{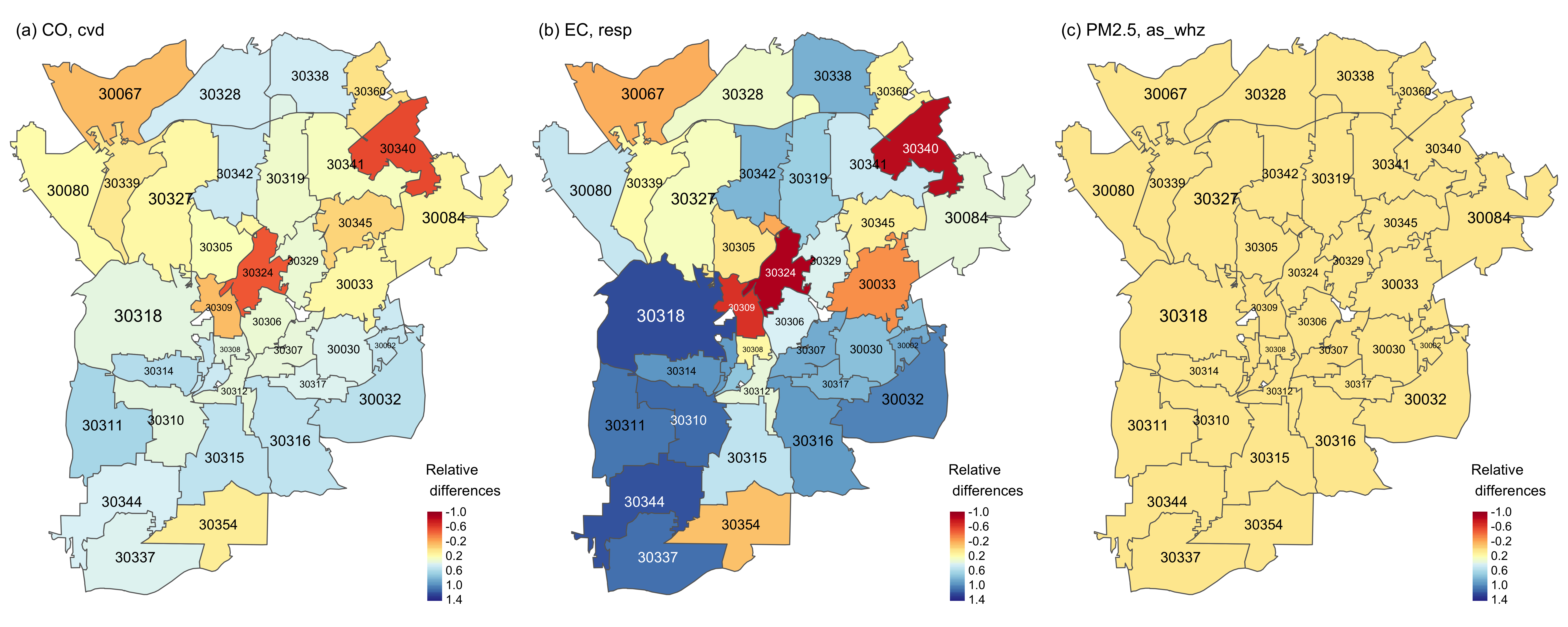}
    \caption{Relative differences of the ZCTA-level number of exposure-attributable ED visits between using the proposed exposure-quantile model and the conventional mean-exposure model (as\_whz = asthma or wheeze ED visits, resp = respiratory disease ED visits, cvd = cardiovascular disease ED visits).}
    \label{fig:mapAD}
\end{figure}

For short-term associations between ambient concentrations of CO and respiratory ED visits and cardiovascular disease ED visits, as shown in Figures \ref{fig:SHEDSbetatau}(b) and (c), the proposed model including quantile functions was preferred, and we found that effects of CO at lower quantile levels are more pronounced. The corresponding short-term associations for the two selected contrasts are presented in Figure \ref{fig:SHEDSRR}(e). We observed that simply using average concentrations of CO as the covariate in the health model underestimated relative risk of CO on cardiovascular disease ED visits and respiratory disease ED visits compared with the proposed model that considers the entire exposure distributions. Differences in the estimated number of ED visits attributed to the exposure to air pollution are also present. For example, Figures \ref{fig:mapAD}(a)-(b) illustrates differences in the number of exposure-attributable ED visits by ZCTA when effects of air pollutants vary by their quantile levels. Specifically, the total number of cardiovascular ED visits attributed to CO exposure was estimated to be 798 (95\% CI: 74 - 3,240) using the proposed model, while the mean model yielded an estimate of 640 (95\% CI: -80 - 1,343). In contrast, for PM$_{2.5}$ and asthma or wheeze ED visits where health effects are invariant with respect to quantile levels, the proposed model and the conventional model led to similar results (Figure \ref{fig:mapAD}(c)). 

\section{Discussion}
In this work, we propose a scalar-on-quantile-function approach to fully characterize effects of environmental exposures on aggregate health outcomes by treating exposure quantile functions as functional covariates. Compared to methods which solely include summary statistics of personal exposures as the exposure metric of interest, our approach accounts for within-group exposure heterogeneity and allows more flexible associations between exposures and aggregate health outcomes. In addition, parametric distribution assumptions on exposure distributions are not necessary in our approach. With the proposed Bayesian two-stage estimation procedure, estimates of health effects can be obtained while incorporating uncertainties in the estimation of exposure quantile functions. This two-stage procedure also alleviates the computation burden when associations between one exposure and multiple health outcomes are examined, compared to approaches that jointly model exposure and health data.  

Applying the proposed model to the motivating ED visits data in Atlanta, we identified novel short-term associations between ambient air pollution concentrations and ED visits which were masked when daily population average exposures were linked to ED visits. For example, results suggest that effects of ambient concentrations of CO on respiratory disease ED visits and cardiovascular disease ED visits are highest at lower quantile levels. These new findings may be important for identifying subpopulations most vulnerable to ambient air pollution. For the majority of pollutant and outcome pairs, we found robust and positive associations comparable to the conventional approach, which strengthens prior evidence on the negative health effects of air pollution.  

The proposed model was motivated by the SHEDS data which provide simulated personal exposure to different air pollutants in Atlanta over four years. While personal exposures are not widely available in large scale population-based epidemiological studies, the proposed model can be applied to more general settings. Specifically, the proposed approach is applicable for scenarios where within group exposure heterogeneity exists and can be characterized using quantile functions. For example, in the scenario where environmental exposures are predicted using spatial-temporal models at finer spatial resolutions compared to spatial resolution of the health data, our proposed approach can be applied with exposure quantile functions derived from predicted exposures and population density. In addition, the health outcome does not have to to be restricted to aggregated counts. For example, in the application of analyzing birthweight data in North Carolina conducted by \cite{berrocal2011use}, the effects of an individual's exposure distribution can be assessed using our approach. 

In this work, we focus on single environmental exposure. One further extension of the proposed approach is to simultaneously examine effects of multiple exposures on aggregate health outcome. A possible strategy is to incorporate quantile surface of exposures as the functional predictor in the health model, which might account for correlations between exposures. In the real data analysis, we found that the estimation of exposure quantile functions is sensitive to outliers. As a result, larger uncertainties of estimated quantile functions are observed in the distribution tails and the estimation of health effects may be affected. To mitigate the impact of outliers on the exposure quantile estimation, one could consider introducing parametric methods for characterizing tails of exposure distributions \citep{zhou2012estimating}.

\section{Acknowledgements}
We thank Lisa Baxter and Kathie Dionisio from the US EPA for providing the SHEDS exposure data. 

\bibliographystyle{apalike}
\bibliography{ref}

\section{Funding}
The work is supported by grant R01ES027892 and R01ES028346 from the National Institutes of Environmental Health. The content is solely the responsibility of the authors and does not necessarily represent the official views of the National Institutes of Health.

\end{document}


\maketitle

This supplementary document includes the basis functions introduced in Section 3.3.2 used for expanding exposure quantile functions, details of MCMC algorithms in Section 4, and additional results of percent increase in risk reported in Section 5.2.

\renewcommand{\thesection}{S\arabic{section}}

\setcounter{equation}{0}
\renewcommand{\theequation}{S\arabic{equation}}

\setcounter{table}{0}
\renewcommand{\thetable}{S\arabic{table}}

\setcounter{figure}{0}
\renewcommand\thefigure{S\arabic{figure}}  

\begin{singlespace}
\begin{table}[H]
\small
\begin{center}
\caption{Estimates of percent (\%) increase in risk associated with one unit shift to right for the distribution of air pollutant concentrations from analyzing the motivating data.}
\label{t:RRincrease_SHEDS}
\setlength\tabcolsep{2pt}
\renewcommand{\arraystretch}{1.5}
\begin{tabular}[1.2\textwidth]{cc cc c cc}
\hline \hline 
 &  & \multicolumn{2}{c}{using estimated quantile functions} & & \multicolumn{2}{c}{using sample mean of individual exposures} \\ \cline{3-4} \cline{6-7} 
 air pollutant & ED visit $^a$ &  estimate & 95\% CI & & estimate & 95\% CI \\ \hline
 \multirow{3}{*}{CO} & asthma/wheeze & 2.84 & (0.09, 5.67) && 1.44 & (-0.09, 3.00) \\
 & resp & 2.42 & (0.79, 4.06) && 0.73 & (-0.15, 1.63) \\
 & cvd & 3.29 & (0.53, 6.16) && 1.40 & (-0.17, 2.96) \\ \hline
 \multirow{3}{*}{EC} & asthma/wheeze & 2.80 & (-0.17, 5.80) && 1.39 & (0.14, 2.65)) \\
 & resp & 2.58 & (0.82, 4.31) && 0.56 & (-0.14, 1.26) \\
 & cvd & 2.26 & (-0.45, 5.09) && 1.12 & (-0.16, 2.41) \\ \hline
\multirow{3}{*}{NO$_\text{x}$} & asthma/wheeze & 135.51 & (-48.37, 965.89) && 42.00 & (-14.03, 132.80) \\
 & resp & 1.78 & (-58.17, 144.89) && 0.46 & (-24.49, 33.50) \\
 & cvd & -28.37 & (-84.01, 206.99) && 24.07 & (-23.92, 104.32) \\ \hline
 \multirow{3}{*}{PM$_{2.5}$} & asthma/wheeze & 0.83 & (0.09, 1.60) && 0.82 & (0.53, 1.10) \\
& resp & 0.47 & (0.02, 0.94) && 0.51 & (0.34, 0.68) \\
& cvd & -0.07 & (-0.81, 0.66) && 0.42 & (0.12, 0.72) \\ \hline \hline
\end{tabular}
\end{center}\vspace*{-10pt}
{$^a$ \footnotesize resp = respiratory disease, cvd = cardiovascular disease.} \\ 
\end{table}
\end{singlespace}


\section{Basis functions for modeling exposure quantile functions} \label{app:basisfun_quan}
Let $0=\kappa_1 < \dots < \kappa_{L+1}=1$ denote two boundaries and $L-1$  equally spaced inner knots over the interval $[0,1]$. The $l$-th basis function $B_l(\tau)$ with $\kappa_l < 0.5$ is defined as:
\begin{equation*}
    B_l(\tau) = [ F^{-1}(\kappa_l) -  F^{-1}(\kappa_{l+1}) ] \boldsymbol{I}(\tau < \kappa_l) + [F^{-1}(\tau) -  F^{-1}(\kappa_{l+1})]\boldsymbol{I}(\kappa_l \le \tau < \kappa_{l+1}) + 0 \boldsymbol{I}(\kappa_{l+1} \le \tau),
\end{equation*}
for the $l$-th basis function with $\kappa_l \ge 0.5$, $B_l(\tau)$ is defined as:
\begin{equation*}
    B_l(\tau) = 0 \boldsymbol{I}(\tau < \kappa_l) + [ F^{-1}(\tau) -  F^{-1}(\kappa_{l})]\boldsymbol{I}(\kappa_l \le \tau < \kappa_{l+1}) + [F^{-1}(\kappa_{l+1}) -  F^{-1}(\kappa_{l})] \boldsymbol{I}(\kappa_{l+1} \le \tau),
\end{equation*}
where $F^{-1}$ is the standard normal quantile function for piecewise Gaussian functions or the quantile function of Gamma distribution with pre-specified shape and scale parameters for piecewise Gamma functions, and $\boldsymbol{I}(\cdot)$ is an indicator fucntion. Following \cite{smith2013bsquare}, the shape and scale parameters are chosen to be 5 and 1, respectively. 

\section{MCMC algorithms} \label{app:MCMC}
\subsection{Polya-Gamma distributions}
The following facts about PG distribution has been presented in \cite{pillow2012fully} and \cite{polson2013bayesian}.
\begin{enumerate}
    \item For any choice of $a$ and a PG random variable $\omega \sim PG(b, 0)$, where $b > 0$, we have 
    \begin{equation}
        \frac{\exp{(\eta)}^{a}}{[1 + \exp{(\eta)}]^{b}} = 2^{-b}\exp{(\kappa \eta)}\int_{0}^{\infty} \exp{(-\omega\eta^2/2)}p(\omega|b, 0) d\omega,
    \end{equation}
where $p(\omega|b, 0)$ denotes the density function of random variable $\omega$ and $\kappa = a - b/2$. 
\item The density function of PG random variable $\omega^{*} \sim PG(b, c)$, where $b > 0$ can be written as:
\begin{equation}
    p(\omega^{*} | b, c) = \frac{\exp(-c^2 \omega^{*}/2)p(\omega^{*}|b, 0)}{\int_{0}^{\infty} \exp{(-c^2\omega/2)}p(\omega|b, 0)d\omega}
\end{equation}
\end{enumerate}

\subsection{MCMC algorithm for estimating exposure quantile functions} \label{app:MCMCquan}
Consider a case where exposure quantile functions are correlated across groups by allowing the basis coefficients to follow a first-order Gaussian Markov random field process (the same setting as in Section 4 and in the estimation of quantile functions of NO$_{\text{x}}$ and PM$_{2.5}$ in our application). Specifically, the quantile function for group $i$ is specified as:
\begin{equation*}
  \begin{gathered}
   Q_{i}(\tau) = \theta_{i,0} + \sum_{l=1}^{L}B_{l}(\tau)\theta_{i,l}, \; i=1,\dots,n, \\
   (\theta_{0,1}, \dots, \theta_{0,n})^T \sim MVN(\theta_{0}, \Sigma_0), \; \Sigma_0 = \sigma_{0}^{2}(D_w - \rho_{0}W)^{-1}, \\
    (\theta_{l,1}^{*}, \dots, \theta_{l,n}^{*})^T \sim MVN(\theta_{l}, \Sigma_1), \; \Sigma_1 = \sigma_{1}^{2}(D_w - \rho_{1}W)^{-1} \; \text{for } l = 1,\dots, L, \theta_{l,i} = \texttt{max}\{\theta_{l,i}^{*}, 0.01\},
    \end{gathered}
\end{equation*}
where $W$ is a symmetric adjacency matrix; and $D_{w}$ is a diagonal matrix with the $i$-th diagonal element equal to the row sum of $i$-th row of $W$. Let $x_{ij}$ denote the exposure for $j$-th observation for group $i$, $j=1,\dots,m_{j}$, $i=1,\dots,n$, and let $\boldsymbol{x}_i$ denote the vector of individual exposures collected for group $i$.
\begin{enumerate}
\item Update $(\theta_{0,1},\dots,\theta_{0,n})$ using Metropolis-Hasting (MH) algorithms 
\begin{equation*}
\pi \big(\theta_{0,i} | \{\boldsymbol{x}_{i}\}_{i=1,\dots,n}, \{\theta_{l,i}\}_{l=1,\dots,L, i=1,\dots,n}, \{\theta_{0,i': i' \ne i}\} \big) \propto \prod_{j=1}^{m_i} f_i \big(x_{ij}|\theta_{0,i},\{\theta_{l,i}\}_{l=1,\dots,L} \big) \pi(\theta_{0,i}|\{\theta_{0,i': i' \ne i}\}),
\end{equation*}
where $ f_i \big(x_{ij}|\theta_{0,i},\{\theta_{l,i}\}_{l=1,\dots,L} \big) = \Big[\frac{d Q_i(\tau)}{d \tau}\Big]^{-1} \biggr\rvert_{\tau^*:Q_{i}(\tau^*) = x_{ij}}$, $\pi(\theta_{0,i}|\{\theta_{0,i': i' \ne i}\})$ denotes the conditional normal density function derived by assuming $(\theta_{0,1}, \dots, \theta_{0,n})^T \sim MVN(\theta_{0}, \Sigma_0), \; \Sigma_0 = \sigma_{0}^{2}(D_w - \rho_{0}W)^{-1}$. 


\item Update $(\theta_{l,1}^{*}, \dots, \theta_{l,n}^{*})$ for $l=1,\dots,L$ \\
Follow the similar step as outlined in Step 1 for updating $(\theta_{0,1},\dots,\theta_{0,n})$ but replacing $\pi(\theta_{0,i}|\{\theta_{0,i': i' \ne i}\})$ with $\pi(\theta_{l,i}^{*}|\{\theta^*_{l,i': i' \ne i}\})$, which is the conditional normal density derived from knowing $(\theta_{l,1}^{*}, \dots, \theta_{l,n}^{*})^T \sim MVN(\theta_{l}, \Sigma_1), \; \Sigma_1 = \sigma_{1}^{2}(D_w - \rho_{1}W)^{-1}$. Finally, letting $\theta_{l,i} = \texttt{max}\{\theta_{l,i}^*, 0.01\}$. 

\item Update $\theta_{0}$ \\
Introducing the normal prior $N(0, 100)$ for $\theta_{0}$, the conjugate full conditional distribution of $\theta_{0}$ is $N(a_n, A_n)$
\begin{gather*}
    A_n = (\boldsymbol{X}^T \Sigma_{0}^{-1} \boldsymbol{X} + C^{-1})^{-1}, \\
    a_n = A_n \times \big(\boldsymbol{X}^T \Sigma_{0}^{-1} \boldsymbol{\theta}_{0,\boldsymbol{\cdot}} + C^{-1}c\big),
\end{gather*}
where $\boldsymbol{X}$ is a column vector of 1 with length of $n$, $\boldsymbol{\theta}_{0,\boldsymbol{\cdot}} = (\theta_{0,1}, \dots, \theta_{0, n})^T$, $c$ and $C$ are mean and variance of the normal prior assumed for $\theta_{0}$. 

\item Update $\theta_{l}$ for $l=1,\dots,L$ \\
Follow the same step as outlined for updating $\theta_{0}$, but replacing $\Sigma_{0}^{-1}$ with $\Sigma_{1}^{-1}$, and $\boldsymbol{\theta}_{0,\boldsymbol{\cdot}}$ with $\boldsymbol{\theta}^{*}_{l,\boldsymbol{\cdot}}$.

\item Update $\sigma_0^2$ \\
With Inverse-Gamma(0.1, 0.1) prior for $\sigma_0^2$, the conjugate full conditional distribution is Inverse-Gamma($a_n$, $b_n$)
\begin{gather*}
    a_n = 0.1 + n/2, \\
    b_n = 0.1 + \frac{(\boldsymbol{\theta}_{0,\boldsymbol{\cdot}} - \theta_{0})^T (D_w - \rho_0 W) (\boldsymbol{\theta}_{0,\boldsymbol{\cdot}} - \theta_{0}) }{2}.
\end{gather*}

\item Update $\sigma_1^2$ \\
With Inverse-Gamma(0.1, 0.1) prior for $\sigma_1^2$, the conjugate full conditional distribution is Inverse-Gamma($a_n$, $b_n$)
\begin{gather*}
    a_n = 0.1 + \frac{L\times n}{2}, \\
    b_n = 0.1 + \frac{\boldsymbol{z}_i^T (D_w - \rho_1 W) \boldsymbol{z}_i }{2}, 
\end{gather*}
where $\boldsymbol{z}_i = \big( (\boldsymbol{\theta}^{*}_{1,\boldsymbol{\cdot}} - \theta_{1})^T, \dots, (\boldsymbol{\theta}^{*}_{L,\boldsymbol{\cdot}} - \theta_{L})^T \big)^T$ is a column vector of length $L\times n$, $\boldsymbol{\theta}^{*}_{l,\boldsymbol{\cdot}} = (\theta^{*}_{l,1},\dots,\theta^{*}_{l,n})^T$ for $l=1,\dots,L$.  

\item Update $\rho_0$ \\ 
Since we assign discrete prior for $\rho_0$, the point density of the full conditional distribution is proportional to $p(\boldsymbol{z}_{0} | \sigma^2, \rho_{0}) \pi(\rho_{0})$. The below equations present an efficient computation of the log density of $\boldsymbol{z}_{0}$. 
\begin{equation*}
\begin{split}
 &\log \big( p(\boldsymbol{z}_{0} | \sigma_0^2, \rho_{0}) \pi(\rho_{0}) \big ) \\
 &\propto -\frac{1}{2}\log|\sigma_0^2(D - \rho_{0})^{-1}| - \frac{1}{2 \sigma_0^2}\boldsymbol{z}_{0}^{T}(D - \rho_{0}W)\boldsymbol{z}_{0} \\
 &\propto \frac{1}{2}\log|D - \rho_{0}W| + \frac{\rho_{0}}{2 \sigma_0^2}\boldsymbol{z}_{0}^{T}W \boldsymbol{z}_{0} \\
 &= \frac{1}{2}\log|D(I - \rho_{0}D^{-1}W)| + \frac{\rho_{0}}{2 \sigma_0^2}\boldsymbol{z}_{0}^{T}W \boldsymbol{z}_{0}   \\
 &\propto \frac{1}{2}\log|(I - \rho_{0}D^{-1}W)| + \frac{\rho_{0}}{2 \sigma_0^2}\boldsymbol{z}_{0}^{T}W \boldsymbol{z}_{0} = \frac{1}{2}\sum_{i=1}^{n} \log(1 - \rho_{0} \lambda_{i}) + \frac{\rho_{0}}{2 \sigma_0^2}\boldsymbol{z}_{0}^{T}W \boldsymbol{z}_{0} ,
\end{split}
\end{equation*}
where $\boldsymbol{z}_0 = \boldsymbol{\theta_{0, \boldsymbol{\cdot}}} - \theta_0$, and $\lambda_{i}$ is the $i$-th eigenvalue of $I - \rho_{0}D^{-1}W$. 

\item Update $\rho_1$ \\
Follow the similar step as outlined for updating $\rho_0$.  
\end{enumerate}

\subsection{MCMC algorithm for estimating health effects} \label{app:MCMChealth}
Consider a proposed model that has intercept but does not include other covariates $\boldsymbol{Z}_i$ and the mean-zero residual process $\epsilon_i$:
\begin{equation*}
    \begin{gathered}
   Y_i|\lambda_i \sim Poisson(\lambda_i) \quad \lambda_i|\xi, \eta_i \sim Gamma(\xi, \exp(\eta_i)), \\
   \eta_i = \beta_{0} + \int_0^1\beta(\tau)Q_i(\tau)d\tau, \; i = 1, \dots, n, \\
    \end{gathered}
\end{equation*} The density function, expectation, and variance of $Y_{i}$ can be expressed as:
\begin{gather*}
    p(y_{i}|\xi, \eta_{i}) \propto (1 - q_{i})^{\xi}q_{i}^{y_{i}}, \; \text{where } q_{i} = \frac{\exp(\eta_{i})}{1 + \exp{(\eta_{i}})} \\
    E(Y_{i}) = \frac{q_{i}\xi}{1 - q_{i}} = \xi \exp(\eta_{i}) = \mu_{i} \\
    Var(Y_{i}) = \frac{q_{i}\xi}{(1-q_{i})^2} = \xi \exp{\eta_{i}}(1 + \exp{\eta_{i}}) = \mu_{i} + \frac{1}{\xi} \mu_{i}^2
\end{gather*}
Following \cite{pillow2012fully}, a latent PG random variable is introduced to enable the implementation of Gibbs sampling for full Bayesian inference of over-dispersed log-linear models. Let $\omega_{i}$ denote the latent variable following $PG(y_{i} + \xi, \eta_{i})$. 

\subsubsection{exposure quantile functions are known} \label{app:MCMChealth_known}
Let $\boldsymbol{X}_{i}^* = \int_{0}^{1} \boldsymbol{K}(\tau) Q_i(\tau) d\tau$, where $\boldsymbol{K}(\tau) = \big( K_{0,p}(\tau), \dots, K_{p,p}(\tau) \big)^T$ is a vector of orthonormal Bernstein polynomials of degree $p$, and $\beta(\tau) = \sum_{j=0}^p K_{j,p} (\tau) \beta_{j}$ (refer to Eqn. (3) in the manuscript for more details). 
\begin{enumerate}
\item Update latent variable $\omega_{i}$ \\
$\omega_{i} | y_{i}, \eta_{i}, \xi \sim PG(y_{i} + \xi, \eta_{i})$, draw $\omega_{i}$ for each observation using \texttt{R} function \texttt{rpg} in package \texttt{BayesLogit}.  
\item Update $\boldsymbol{\beta} = (\beta_0, \beta_{1}, \dots, \beta_{p})^T$ \\
Given $\boldsymbol{y} = (y_{1}, \dots, y_{n})$, $\boldsymbol{\omega} = (\omega_{1}, \dots, \omega_{n})$, $\boldsymbol{\eta} = (\eta_{1}, \dots, \eta_{n})$, the full conditional distribution of $\boldsymbol{\beta}$ is given by   
\begin{equation*}
\begin{split}
& \pi(\boldsymbol{\beta} | \boldsymbol{y}, \boldsymbol{\eta}, \boldsymbol{\omega}, \xi) \\
&\propto \pi(\boldsymbol{\beta}) \times \prod_{i=1}^{n} \frac{\exp(\eta_{i})^{y_{i}}}{[1 + \exp(\eta_{i})]^{(y_{i}+\xi)}}p(\omega_{i}| y_{i} + \xi, \eta_{i}) \\
& \stackrel{\text{Eqn. (S1)}}{\propto} \pi({\boldsymbol{\beta}}) \prod_{i=1}^{n} \exp{\{\frac{1}{2}(y_{i}-\xi)\eta_{i}\}} \int_{0}^{\infty} e^{-\omega_{i}\eta_{i}^2/2} p(\omega_{i}|y_{i}+\xi, 0) d\omega_{i} \times p(\omega_{i}| y_{i}+\xi, \eta_{i}) \\
& \stackrel{\text{Eqn. (S2)}}{=} \pi({\boldsymbol{\beta}}) \prod_{i=1}^{n}  \exp{\{\frac{1}{2}(y_{i}-\xi)\eta_{i}\}} \frac{e^{-\omega_{i}\eta_{i}^2/2} p(\omega_{i}|y_{i}+\xi, 0)}{p(\omega_{i}| y_{i} + \xi, \eta_{i})} \times p(\omega_{i}| y_{i} + \xi, \eta_{i}) \\
&= \pi({\boldsymbol{\beta}}) \prod_{i=1}^{n} \exp{\{\frac{1}{2}(y_{i}-\xi)\eta_{i}\}} \exp{\{-\omega_{i}\eta_{i}^2/2\}} p(\omega_{i}|y_{i} + \xi, 0) \\ 
& \propto \pi({\boldsymbol{\beta}}) \prod_{i=1}^{n} \exp{ \big\{ -\frac{\omega_{i}}{2} \big( \frac{y_{i}-\xi}{2\omega_{i}} - \eta_{i}\big)^2 \big\} } 
\end{split}
\end{equation*}
Given the prior of $\boldsymbol{\beta}$ is $MVN(\boldsymbol{c}, C)$, where $ \boldsymbol{c} = \boldsymbol{0}$ and $C = \text{diag}(100)$. The conjugate full conditional distribution of $\boldsymbol{\beta}$ is $MVN(\boldsymbol{a}_n, A_n)$, which is given by 
\begin{gather*}
A_{n} = \big( \boldsymbol{X}^{T}\Omega\boldsymbol{X} + C^{-1} \big)^{-1}, \\ 
\boldsymbol{a}_{n} = A_{n} \big(C^{-1}\boldsymbol{c} + \boldsymbol{X}^{T} \Omega \boldsymbol{z} \big),
\end{gather*}
where $\Omega = \text{diag}(\boldsymbol{\omega})$ is a $n \times n$ square matrix, $\boldsymbol{X}$ is a design matrix corresponding the parameter vector $\boldsymbol{\beta}$ with size of $n \times (p+1)$. We also define $z_{i} = \frac{y_{i} - \xi}{2\omega_{i}}$ and $\boldsymbol{z} = (z_{1}, \dots, z_{n})$ ($n \times 1$ vector). This Gibbs sampling algorithm can be applied for models that include $\boldsymbol{Z}_i$ and $\epsilon_i$ by modifying the design matrix. 

\item Update $\xi$ \\
As suggested in \cite{neelon2019bayesian}, we choose zero-truncated normal distribution centered at the current value as the proposal distribution, the corresponding variance (i.e., tuning parameter) is denoted as $d_{\xi}$. Given the uniform prior, the full conditional distribution of $\xi$ is:
  \begin{equation*}
\begin{split}
\pi(\xi | \boldsymbol{y}, \boldsymbol{\eta}) \propto \prod_{i=1}^{n} \frac{\exp(\eta_{i})^{y_{i}}}{\{1 + \exp(\eta_{i})\}^{y_{i} + \xi}} \pi(\xi)  
\end{split}
\end{equation*}
\end{enumerate}

\subsubsection{exposure quantile functions are estimated} \label{app:MCMChealth_est}
To propagate uncertainties associated with estimating exposure quantile functions, MVN prior is assumed for $\boldsymbol{\theta}_{\boldsymbol{\cdot}, i} = (\theta_{0,i}, \theta_{1, i}, \dots, \theta_{L,i})$, $L=4$ in our simulation studies and application. With $MVN(\hat{\boldsymbol{\theta}}_{\boldsymbol{\cdot}, i}, \Lambda_i)$ prior introduced for $\boldsymbol{\theta}_{\boldsymbol{\cdot}, i}$, this parameter vector can be updated when fitting the health model with its conjugate full conditional distribution following a MVN distribution, where $\hat{\boldsymbol{\theta}}_{\boldsymbol{\cdot}, i}$ and $\Lambda_i$ are computed from its posterior predictive distribution obtained from estimating exposure quantile functions. For updating other parameters (e.g., $\boldsymbol{\beta}$, $\xi$), following same procedures outlined in Section \ref{app:MCMChealth_known} of the supplementary materials.

\bibliographystyle{apalike}
\bibliography{ref}